\documentclass[a4paper,fleqn]{cas-sc}

\usepackage[numbers]{natbib} 

\newcommand{\systemname}{\textsc{Upstanders' Practicum}\xspace}
\newcommand{\papertitle}{Attention: What Prevents Young Adults from Speaking Up Against Cyberbullying in an LLM-Powered Social Media Simulation} %

\usepackage{longtable}
\usepackage{caption} 
\usepackage{enumitem}

\usepackage{siunitx}
\usepackage{tabularx}
\usepackage{bbding}
\usepackage{longtable}
\usepackage{ifthen}

\usepackage{tcolorbox}
\newtcolorbox{promptbox}{
  colback=gray!5, colframe=gray!30,
  boxrule=0.4pt, arc=2pt,
  left=6pt, right=6pt, top=4pt, bottom=4pt,
  fontupper=\small
}

\usepackage[normalem]{ulem}

\newcommand{\deletes}[1]{}

\usepackage[backgroundcolor=white, textcolor=magenta, linecolor=magenta, bordercolor=white, textsize=tiny]{todonotes}

\newcommand{\placeholder}[1]{\textcolor{gray}{#1}}
\usepackage{lipsum}
\newcommand{\placeholderPARA}[1][]{
  \ifthenelse{\equal{#1}{}}{\placeholder{\lipsum[2]}}{\placeholder{\lipsum[1][1-#1]}}
}
\newcommand{\todocite}[1][]{
  \ifthenelse{\equal{#1}{}}{\textcolor{magenta}{[REF]}\xspace}{\placeholder{[#1]\xspace}}
}

\def\tsc#1{\csdef{#1}{\textsc{\lowercase{#1}}\xspace}}
\tsc{WGM}
\tsc{QE}

\begin{document}
\let\WriteBookmarks\relax
\def\floatpagepagefraction{1}
\def\textpagefraction{.001}

\shorttitle{}    

\shortauthors{}  

\title [mode = title]{\papertitle}



\author[1]{Qian Yang}[orcid=0000-0002-3548-2535]
\cormark[1]
\ead{qianyang@cornell.edu}
\credit{Funding acquisition, Conceptualization, Methodology, Supervision, Writing -- original draft, Writing -- review \& editing}

\author[1]{Jessie Jia}[orcid=0009-0009-2879-5849]
\ead{hj359@cornell.edu}
\credit{Software, Investigation, Data curation, Formal analysis}

\author[1]{Elaine Tsai}
\ead{ect65@cornell.edu}
\credit{Software, Investigation, Data curation, Formal analysis}

\author[1]{Amy Li}
\ead{ayl53@cornell.edu}
\credit{Investigation, Data curation, Formal analysis}

\author[1]{Nader Akoury}
\ead{na476@cornell.edu}
\credit{Software, Supervision}

\author[1]{Natalie N. Bazarova}[orcid=0000-0001-5375-6598]
\ead{bazarova@cornell.edu}
\credit{Funding acquisition}

\affiliation[1]{organization={Cornell University},
            city={Ithaca},
            state={New York},
            country={USA}}

\cortext[1]{Corresponding author}


\begin{abstract} 
Interactive, multi-agent social simulation systems have shown promise for helping users practice navigating various complex social situations across domains.
This paper asks: To what extent can such systems help young adult (YA) bystanders speak up publicly against cyberbullying, a task often thwarted by complex, multi-party social dynamics?
We created \textsc{Upstanders' Practicum}, a multi-AI-agent social media simulation powered by Large Language Models (LLMs), as a probe and observed 34 YAs freely practicing public bystander intervention across three iteratively refined versions.
We found that practicing public bystander intervention in the simulation was helpful, \textit{but after participants made three attention shifts}: (1) from inattention to paying true attention, (2) from self-focus (``\textit{I don't usually do this''}) to attending to those directly involved, and (3) from resolving the private conflict between bully and victim (``\textit{maybe I could set up the meeting between them''}) to addressing the broader audience online (``\textit{public comment is about norm-setting''}).
Only after these shifts did practice in the simulation start to help: participants then saw a reason to speak up publicly and, through continued practice, crafted tactful public messages without explicit instruction.
These findings illuminate new design and research opportunities for bystander education beyond social skill instruction, namely, \textit{designing for true attention}, \textit{for fostering a vocal upstander identity}, and \textit{for seeing bystander intervention as public norm setting}.
In addition, we open-source \textsc{Truman Agents} (\url{cornell-design-ai-group.github.io/TrumanAgents/}), the first-of-its-kind multi-LLM-agent social media simulation platform that \textsc{Upstanders' Practicum} builds upon, for future cyberbullying and social media research.

\end{abstract}



\begin{highlights}
\item LLM simulations train bystanders' social skills to speak up against cyberbullying.
\item We observed 34 YAs practicing public intervention in a multi-agent LLM simulation
\item Attention shifts (inattention, self, others, online norms) preceded intervention.
\item Once attention shifted, YAs spoke up tactfully against cyberbullying on their own.
\item LLM bystander training should target inattention and lurker identity, not just skills.
\end{highlights}

\begin{keywords}
Cyberbullying \sep Artificial Intelligence \sep Large Language Models 
\sep Social Simulation \sep Bystander Intervention \sep Human-AI Interaction
\end{keywords}

\maketitle
\section{Introduction}

Cyberbullying on social media causes serious harm, yet young adults (YAs), the largest demographic of social media users, rarely speak up when they witness it~\cite{pew2017-experience-online-harassment,davidovic2023intervene,song2018factors}.
This silence is costly: when bystanders stay quiet, victims feel isolated, perpetrators feel emboldened to escalate, and passive observers come to see abuse as normal, potentially creating a downward spiral of toxicity online~\cite{garland2022impact,hawkins1998naturalistic,Aleksandric-bystander-first-response-www23}.

Interactive technologies hold great promises for helping YA bystanders speak up against cyberbullying~\cite{difranzo2018upstanding,garland2022impact}.
Speaking up publicly does what private support cannot: it sets visible prosocial norms for the entire community. Research on norm perception shows that people calibrate their own behavior to what they see others do~\cite{tankard2016norm}; a single visible act of speaking up therefore reaches not just those directly involved but present and future passive observers~\cite{dominguez2018systematic,xiao2022AdolescentsNeeds}.
Acute teacher shortages in digital literacy and online safety~\cite{dee2017teacherShortage,TeacherShortageAreas} mean in-person training alone cannot prepare enough YAs to take on this role.
Within such a milieu, systems that help YAs practice public intervention at scale could transform not only individual responses to abuse, but online culture at large.

Recent work on multi-agent social simulation offers a particularly promising instantiation of this vision~\cite{YangSocialSkillTraining2024_Arxiv}.
Because complex social skills require practice in realistically complex social situations~\cite{kass1994constructing,holt2017multidisciplinary}, multi-agent simulations are uniquely suited: users interact with multiple Large Language Model (LLM) agents who assume distinct roles, experiment with various ways of navigating multi-party social dynamics, and learn from the social consequences of their actions. From conflict resolution to consensus-finding, this approach has proven effective across a range of social skill training contexts~\cite{ShaikhCHI24_rehearsal,AGORA_PradeepCHI26EA,GLOSS_guevarraAAAI2025,roleplayDoh_LouieEMNLP2024}.

We ask: \emph{To what extent can this approach help YA bystanders speak up publicly against cyberbullying, a task often thwarted by complex, multi-party social dynamics~\cite{song2018factors,casey2017situational}?}
Most YAs already recognize cyberbullying as wrong and want to help~\cite{davidovic2023intervene}; what holds them back is low confidence in their ability to read the social dynamics and respond tactfully without escalating the situation~\cite{jenkins2019bystander, metallidou2018bystanders, yule2020college}. Open-ended, scaffolded practice enabled by multi-agent simulation appears well-suited to address these barriers, making it a natural starting point for investigation.

We investigated this question via an iterative, probe-based approach.
We began by creating \systemname, a multi-agent social media simulation with cyberbullying scenarios and characters grounded in prior empirical research, and observing YAs freely practicing public bystander intervention with no support.
As we identified where they struggled, we added support and observed again, refining the system across three versions (N=6+8+20).
We then analyzed participants' actions and reasoning across these iterations.
This iterative approach allowed us to discover to what extent and in what ways multi-agent social simulation helps YAs speak up publicly against cyberbullying.

We found that practicing public bystander intervention in the simulation was helpful, but after participants made three attention shifts that the simulation did not address. By "attention shift," we mean a change in what participants attended to, what occupied their cognitive engagement, as they encountered and responded to cyberbullying in the simulation.
\begin{itemize}[itemsep=0pt,leftmargin=0.5cm]
    \item \textit{From inattention to true attention:} Some participants repeatedly scrolled right past cyberbullying incidents they were fully capable of recognizing; others confused the victim with the bully for up to 30 minutes, even while directly messaging them. Only after truly attending to the situation could they begin to respond.
    \item \textit{From self-focus to those directly involved:} Only after shifting attention from themselves ("\textit{I don't usually do this}") to the victim and bully ("\textit{[I am thinking] how they might want me to respond}") could participants take any action. Their first actions were almost always a direct message to the victim, then to the bully.
    \item \textit{From those directly involved to the broader audience:} Only after shifting attention from the private conflict between bully and victim (``\textit{maybe I could set up the meeting between them}") to the parasocial audience ("\textit{public comment is about norm-setting}", ``\textit{it was more for the audience watching}'') did participants see a reason to speak up publicly. 
\end{itemize}

Among all 34 participants, these shifts always occurred in the same order; every participant who completed all three went on to navigate the social complexities of cyberbullying and speak up through continued practice, without explicit skill instruction.

Based on these findings, we discuss that attentional orientation is an important and under-explored aspect of helping YA bystanders speak up against cyberbullying.
We outline the design and research opportunities this discovery reveals, namely, designing bystander training \textit{for true attention}, \textit{for fostering a vocal upstander identity}, and \textit{for seeing bystander intervention as public norm setting}.

This paper makes three contributions to Human-AI Interaction Design research.
First, it offers a rare empirical account of where YA bystanders get stuck when freely practicing public intervention in a multi-AI-agent simulation. These findings (three attention shifts in particular) offer a point of reference for those designing AI systems that train users to navigate complex social situations online.
Second, it reveals previously unknown design and research opportunities for LLM-empowered bystander training beyond social skill instruction.
Third, we open-source \textsc{Truman Agents} (\url{cornell-design-ai-group.github.io/TrumanAgents/}), a multi-LLM-agent social media simulation platform that \systemname builds upon.
First of its kind and configurable without coding, \textsc{Truman Agents} enables multiple human participants to interact with LLM-driven characters on the same social media feed, opening new possibilities for research on how interactive social simulations can foster prosocial behavior online.


\section{Related Work}

Cyberbullying is common on social media~\cite{pew2017-experience-online-harassment,vogels22TeensCyberbullying,irwin2020SchoolCrime} and its consequences can be severe, including depression, self-harm, and suicide~\cite{kiriukhina2019cyberbullying,kowalski2011cyber,machmutow2012peer,price2010cyberbullying,schneider2012cyberbullying}. Yet young adults (YAs), the largest demographic of social media users, rarely speak up when they witness it: fewer than 30\% intervene in any way, and most interventions are passive, such as flagging the post~\cite{davidovic2023intervene,song2018factors,yule2020college}. 
This section reviews what holds YAs back from speaking up, what interactive technologies researchers have developed to help them, and why multi-agent LLM social simulation is a promising (yet untested) approach to helping YA bystanders speak up.

\subsection{Barriers to YA Bystanders Speaking Up Against Cyberbullying}

YAs (age 18 to 34) are the largest demographic of social media users~\cite{pew2025SocialMediaUse}, and whether they speak up publicly when witnessing cyberbullying plays an irreplaceable role in setting prosocial online norms, reaching not only those directly involved but also present and future observers of the incident~\cite{difranzo2018upstanding,dominguez2018systematic}.
Thankfully, most YAs already possess the knowledge needed to speak up against cyberbullying: Unlike children or teens, they recognize cyberbullying as wrong, empathize with victims, want to help, and know the intervention options ~\cite{song2018factors,davidovic2023intervene,casey2017situational}.
 
Nevertheless, fewer than 30\% of YAs speak up when witnessing cyberbullying~\cite{davidovic2023intervene,yule2020college}. 
Prior work attributes their silence to two interrelated struggles with the complex social dynamics surrounding cyberbullying incidents~\cite{casey2017situational,yule2020college}. 
\begin{enumerate}[leftmargin=0.5cm,itemsep=0pt,topsep=0.3pt,partopsep=0pt]
    \item Uncertainty about what is really going on (``\textit{social cognition}''): cyberbullying situations involve power dynamics and backstories invisible to bystanders, and when bystanders cannot grasp these dynamics, they seldom feel in control enough to act~\cite{Law12ChangingFace,xiao2022AdolescentsNeeds,allison2016cyber,jenkins2019bystander}.
    \item Low confidence in speaking up tactfully (``\textit{social engineering}'') without escalating the situation, causing retaliation, losing peer acceptance, or damaging their own social status~\cite{beckford2015bullies,bastiaensens2015can}.
\end{enumerate}

When bystanders cannot resolve these struggles, they stay silent~\cite{jenkins2019bystander,song2018factors}.

\subsection{Interactive Technologies for Bystander Training}

Interactive technologies hold great promise for helping bystanders intervene against cyberbullying~\cite{difranzo2018upstanding,garland2022impact}. Researchers and educators who build such systems often draw on the Bystander Intervention Model (BIM)~\cite{latane1970unresponsive}, which describes five sequential steps a bystander must complete before taking action: (1) notice the event, (2) interpret it as an emergency, (3) accept personal responsibility, (4) decide how to help, and (5) intervene~\cite{darley1968bystander}.
 
Most bystander training targeted children and adolescents, addressing the barriers they face. Educational programs such as STAC~\cite{midgett2018rethinking_STAC,ueda2021cyberbullying} and systems like \textsc{FearNot!}~\cite{aylett2005fearnot} teach middle schoolers to recognize cyberbullying and empathize with victims, targeting the first two BIM steps.
Interactive systems target later steps. For example, \textsc{Friendly Attac}, a serious game, presents scripted cyberbullying incidents with predefined response options~\cite{desmet2018efficacy}; more recently, researchers have proposed using LLMs to enable open-ended conversation in such role-play systems~\cite{YangGangCHI24_Playwright}. Along similar lines, researchers have proposed and built AI chatbots that provide step-by-step guidance conversationally~\cite{gabrielli2020chatbot,piccolo2021chatbots,YangGangCHI26_CollabUpstanding}. These technologies target bystanders' knowledge and empathy gaps, not the social situational struggles that hold most YAs back.

Moreover, no work has investigated what prevents YAs from speaking up \textit{publicly} against cyberbullying or built systems to help them do so. This gap is notable because researchers have identified public intervention as the most impactful form of bystander action: unlike private support or flagging, a public response can shape the trajectory of the entire conversation that follows~\cite{Aleksandric-bystander-first-response-www23}, and signals to the wider community that abuse is unacceptable~\cite{difranzo2018upstanding,dominguez2018systematic}. Yet the design interventions tested so far target bystander intervention broadly, and the gains observed have been in indirect actions such as flagging cyberbullying posts~\cite{difranzo2018upstanding,taylor2019accountability}.
 
\subsection{LLM Social Simulation for Social Skill Training}
\label{RW3}

Multi-agent LLM simulation has emerged as a promising approach to social skill training, because it lets users practice navigating complex social dynamics with multiple agents who assume distinct roles~\cite{YangSocialSkillTraining2024_Arxiv}.
The approach has proven effective in helping people overcome various social situational challenges, including conflict resolution~\cite{ShaikhCHI24_rehearsal}, civic consensus-finding~\cite{AGORA_PradeepCHI26EA}, structured social skill tutoring~\cite{GLOSS_guevarraAAAI2025}, counseling~\cite{roleplayDoh_LouieEMNLP2024}, and more~\cite{simulating_sreedharCaiIUI2025}.

Interestingly, no work has applied this approach to help YA cyberbullying bystanders speak up publicly against cyberbullying, even though this task is often thwarted by precisely the kind of social situational challenges that scaffolded multi-agent simulation has proven effective at addressing~\cite{song2018factors,casey2017situational,jenkins2019bystander,metallidou2018bystanders,yule2020college}.

Moreover, no tool exists to build such a simulation for social media contexts. Existing multi-agent social media simulations run autonomously and do not support human participants interacting with the simulation or receiving realistic LLM agent responses to their actions~\cite{park2022social,mou-etal-2024-unveiling,zhousotopia,gao2023s3}. This paper addresses both gaps: we first build the tool, then use it to answer our research question.


\section{Method}

This paper asks: \textit{To what extent can multi-agent LLM social simulation help YA bystanders speak up publicly against cyberbullying, a task often thwarted by complex, multi-party social dynamics?} We investigated this question through an iterative, probe-based approach: we refined the system across three versions (N=6+8+20) until we could observe their complete decision-making trajectories, from encountering cyberbullying through to speaking up publicly.

We chose this approach deliberately:

\begin{itemize}[itemsep=0pt,leftmargin=0.5cm]
    \item We chose probes rather than controlled experiments because this is a largely unexplored design space with no established variables to control. No prior work has observed YAs freely practicing bystander intervention in multi-agent simulations (\S\ref{RW3}), so the barriers they encounter are unknown. Probes are well suited to such open-ended inquiry because they surface phenomena that predefined variables might miss. 
    \item We chose an iterative approach rather than a single-version study because upstream barriers can make downstream ones invisible. For example, if participants fail to engage with cyberbullying content at all, no amount of additional observation will reveal what happens after they engage. Only by redesigning to address the upstream barrier and observing new participants can the next layer of barriers come into view.
    \item We are acutely aware that simulation cannot replicate the stakes and bystander behaviors of real social media, and that this poses a risk to the generalizability of our findings. We chose simulation nonetheless, because it allows us to ensure a sufficient number of participants encountered cyberbullying, while controlling the potential harm they experience from it. Importantly, we mitigate the generalizability risk by focusing our analysis on where participants got stuck rather than on what helped them succeed: because simulation is lower-stakes than real social media, barriers that blocked action here would only be stronger in authentic contexts, whereas scaffolding that helped here may not help without it.
\end{itemize}

In what follows, we describe what we built (\S\ref{method1}), who used it (\S\ref{method2}), how we iterated the design based on what we observed (\S\ref{method3}), and how we analyzed the results (\S\ref{method4}).

\subsection{Designing and Implementing the Initial Probe}
\label{method1}

To investigate this question, we set out to build a social media simulation where YAs could freely practice responding to cyberbullying and observe how multiple AI characters responded. We built on \textsc{Truman}~\cite{difranzo2018truman}, an open-source social media simulation platform that provides an Instagram-like interface with core features (posts, comments, likes, direct messages), and extended it with LLM agents that role-play bullies, victims, and bystanders. Participants could post public comments, send direct messages to any agent, browse profiles to investigate backstories, or choose to do nothing.

To ensure participants experienced realistic interactions, we grounded the simulation design in empirical research of real-world cyberbullying in three ways.

\begin{enumerate}[itemsep=0pt,leftmargin=0.5cm]
    \item \textit{Realistic characters.}
    We defined each LLM agent by motives, vulnerabilities, and backstories, basing them strictly on what prior social media research identifies as characteristics that make individuals prone to becoming bullies, victims, or various types of bystanders~\cite{varjas2010high,hamuddin2022they,wilton2011exploration,lipkins2006preventing,Smokowski2019}. 
    For example, in the hazing scenario, the bully holds a position of power and hazes newcomers to assert dominance, enjoy the attention, and carry on a tradition he himself endured~\cite{lipkins2006preventing,wilton2011exploration}; the victim craves social belonging and feels intense pressure to participate rather than resist, for fear of being excluded~\cite{Smokowski2019}. Some bystander agents know the backstory of the bully or victim; others do not. Appendix~\ref{appendix_agent_prompts} provides the full prompts of all LLM agents.

    \item \textit{Realistic situations that create nuanced social dynamics for bystanders to navigate.~}
    For each cyberbullying post, we drew on empirical research to design contexts (including both what is visible or invisible on the social media feed) where these characters' tensions would plausibly escalate into cyberbullying rather than other forms of online conflict or aggression~\cite{boulton1992bully,strawhun2013assessment}.
    For example, in the reckless doxxing scenario, a student posts an embarrassing photo of a friend after a party, genuinely unaware that what they did constitutes cyberbullying~\cite{macallister2016doxing}; the victim withdraws socially, afraid of what else others might share~\cite{franz2023doxing}; and because the bully's intent appears harmless, other bystanders dismiss the situation as a joke, leaving the participant to determine whether intervention is even warranted. Appendix~\ref{appendix_scenario_design} details the full scenario design procedure.

    \item \textit{Realistic feedback on participants' public bystander intervention.~}
    We designed all agents to respond to participant interventions based on their literature-grounded psychological needs, so that participants receive realistic social feedback on their approach: (1) Bully agents apologize or delete their posts only when participants address the underlying needs that prior research identifies as driving their bullying behavior. (2) Victim agents react based on their own needs: they may express gratitude when a participant respects their agency, or distress when an intervention threatens to escalate the situation~\cite{Smokowski2019,franz2023doxing}, and (3) some bystander agents may turn hostile if provoked by insensitive interventions.
    Participants must therefore read the full social dynamics of the situation and respond tactfully: addressing the bully's needs without threatening the victim's sense of agency or arousing passive bystanders into perpetrators. 
    For example, in the reckless doxxing scenario, the bully deletes the post only when a participant both explains that the action constitutes doxxing and acknowledges that it was not the bully's intention to cause harm~\cite{macallister2016doxing}; accusing the bully of intentional bullying causes them to post another embarrassing photo, which in turn deepens the victim's distress and social withdrawal.

    These feedback mechanisms allow participants to experiment with various interventions and experience their respective social consequences, the core mechanism through which social simulation supports complex social skill training~\cite{kass1994constructing,holt2017multidisciplinary}.
\end{enumerate}

We searched the empirical literature extensively for common cyberbullying scenarios that have all the details outlined above. This process yielded four scenario types with rich empirical grounding: intentional hazing, cyberstalking, reckless doxxing, and intentional doxxing.

\subsection{Running User Studies}
\label{method2}

\begin{table}[h]
    \centering
    \renewcommand{\arraystretch}{1.2} 
    \resizebox{\linewidth}{!}{%
    \begin{tabular}{p{0.45\linewidth} | p{0.15\linewidth} p{0.2\linewidth} p{0.2\linewidth}}
    \toprule[2pt]
       & \textbf{Probe 1} (N=6) & \textbf{Probe 2} (N=8) & \textbf{Probe 3} (N=20) \\
       \hline
       \rowcolor{lightgray!20!}
       \multicolumn{4}{c}{\textbf{Age}} \\
       \hline
       18--24 & 3 & 4 & 11 \\ 
       25--34 & 3 & 4 & 9 \\ 
       \hline
       \rowcolor{lightgray!20!}
       \multicolumn{4}{c}{\textbf{Gender}} \\
       \hline
       Women & 3 & 3 & 14 \\
       Men & 3 & 5 & 6 \\
       \hline
       \rowcolor{lightgray!20!}
       \multicolumn{4}{c}{\textbf{Frequency of Social Media Public Posting \& of Speaking Up Against Cyberbullying When It Involves Only Strangers}} \\
       \hline
       Never & 0 \& 3  & 1 \& 7 & 6 \& 16 \\
       Yearly & 1 \& 0 & 2 \& 1 & 6 \& 3 \\
       Monthly & 1 \& 0 & 4 \& 0 & 5 \& 1 \\
       Weekly+ & 1 \& 0 & 2 \& 0 & 3 \& 0 \\
       \bottomrule[2pt]
    \end{tabular}}
    \vspace{0.2cm}
    \caption{Participant demographics across the three probes. We recruited in total $34$ young adults (ages 18--34) with diverse social media posting habits, balancing each probe for demographics and prior posting experience.}
    \label{tab:participants}
\end{table}

\paragraph{Participants.~}
To ensure our findings were not an artifact of recruiting people already comfortable posting publicly, we recruited YAs (ages 18--34) across the full range of posting behaviors, from self-described lurkers who never post to weekly posters. We recruited through social media platforms (Reddit, Prolific) and academic and professional mailing lists, compensating each participant \$$15$/hour. This process resulted in 34 participants (Table~\ref{tab:participants} summarizes demographics, posting behavior, and prior cyberbullying exposure).

Each participant tested only one probe version, with groups balanced for demographics and prior posting experience. Sample sizes varied across probes (N=6, 8, 20): when observations revealed clear patterns, we prioritized redesigning over recruiting additional participants for a version we knew would change. For Probe 3, from which our primary findings derive, the sample size aligns with prior qualitative HCI research~\cite{faulkner2003beyond,sauro2016quantifying}. Throughout the paper, we label participants by probe: P1.1--P1.6 for Probe 1, P2.1--P2.8 for Probe 2, and P3.1--P3.20 for Probe 3.

The study was approved by our institution's IRB. We informed all participants that they would encounter simulated cyberbullying content involving fictional characters, obtained their informed consent, and reminded them they could withdraw at any time without penalty. We monitored for distress throughout each session. In retrospective interviews, we asked whether any content had caused distress; when participants flagged specific concerns, we revised the wording of the relevant posts for subsequent sessions. No participant withdrew from the study.

\paragraph{User Study Procedure:~ Training and Transfer Scenarios.~}
As noted above, a simulation study conducted in a single session cannot fully distinguish genuine cognitive shifts from compliance with scaffolding, which is why our analysis focused on where participants got stuck rather than on what helped them succeed (more on this in~\S\ref{method4}.)

Nevertheless, to partially mitigate this limitation, we created transfer scenarios: a second version of each scenario type with different posts and characters but all scaffolding removed. LLM agents remained present and responsive, but participants received no system-provided support of any kind. This allows participants to practice in familiar scenario types without system support, revealing which barriers persisted despite training.

\paragraph{Data Collection.~}
To understand participants' decision-making trajectories as they practiced speaking up against cyberbullying in the simulation, we collected rich data on both participants' behaviors and thought processes during each session. For behavioral data, we screen-recorded all participant interactions with consent and took field notes, capturing actions such as who participants messaged, what they investigated, and whether and when they posted publicly. Each participant first navigated four training scenarios using their assigned probe (45--60 minutes), then four transfer scenarios (15--20 minutes).

To understand participants' thought processes in detail, we conducted retrospective think-aloud interviews~\cite{YangCHI19_sketchingNLP} immediately after each participant completed all scenarios. Participants watched their screen recordings and narrated their moment-by-moment thinking while we asked clarifying questions. These interviews typically lasted 30--45 minutes. We transcribed all interviews for analysis.


\begin{table}[t] 
  \centering
  \includegraphics[width=\linewidth]{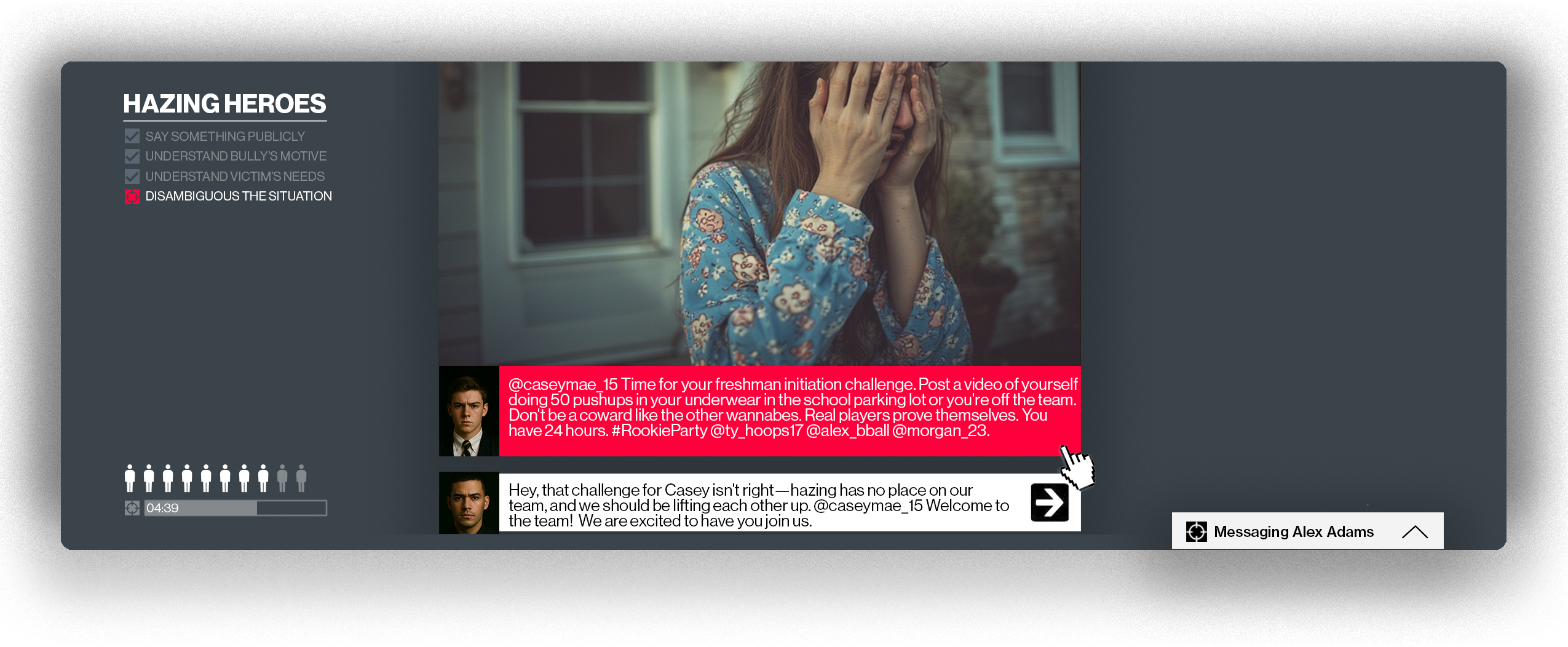}
  \captionof{figure}{Interface design of \systemname (the third and final probe). The simulation presents one cyberbullying incident at a time, with LLM agents playing bullies, victims, and bystanders who respond based on their literature-grounded psychological needs (\S\ref{method1}). A visible checklist (left) specifies \textit{what} to attend to without prescribing \textit{how}, allowing us to observe where participants get stuck when the question of \textit{whether} to intervene is set aside.}
  \label{fig:hero}
\end{table}

\begin{table*}[t]
  \centering
  \footnotesize
  \renewcommand{\arraystretch}{1.4}
  \setlength{\tabcolsep}{4pt}
  \definecolor{dot1}{HTML}{E8A0A0}
  \definecolor{dot2}{HTML}{D35F5F}
  \definecolor{dot3}{HTML}{B22222}
  \definecolor{dot4}{HTML}{6B0F0F}
  \newcommand{\dotI}{\textcolor{dot1}{\scalebox{1.2}{$\bullet$}}\;}
  \newcommand{\dotII}{\textcolor{dot2}{\scalebox{1.6}{$\bullet$}}\;}
  \newcommand{\dotIII}{\textcolor{dot3}{\scalebox{2.1}{$\bullet$}}\;}
  \newcommand{\dotIV}{\textcolor{dot4}{\scalebox{2.7}{$\bullet$}}\;}
  \newcolumntype{R}[1]{>{\raggedright\arraybackslash}p{#1}}
  \begin{tabular}{
    @{}
    R{0.23\linewidth}
    R{0.23\linewidth}
    R{0.25\linewidth}
    R{0.25\linewidth}
    @{}
  }
    \toprule[2pt]
    \textbf{Level 1} \newline \textit{Intentional Hazing}
    & \textbf{Level 2} \newline \textit{Cyberstalking}
    & \textbf{Level 3} \newline \textit{Reckless Doxxing}
    & \textbf{Level 4} \newline \textit{Intentional Doxxing} \\
    \midrule[2pt]

    \rowcolor{lightgray!20}
    \multicolumn{4}{@{}c@{}}{\parbox{\linewidth}{\centering\vspace{3pt}\textbf{Escalating needs for participants' social cognition:}\\[2pt]\textit{The social dynamics becomes increasingly ambiguous} $\longrightarrow$\vspace{3pt}}} \\
    \hline

    \vspace{2pt}
    \dotI \textit{System design:} Bully's motive (e.g., carry on hazing tradition) is visible from the post; victim's situation is relatively transparent from context (e.g., feeling pressure to fit in).
    \vspace{6pt}\newline
    \dotI \textit{Participants can:} act without deep investigation.
    \par\vspace{2pt}

    & \vspace{2pt}
    \dotII \textit{System design:} Bully's motive is visible from the post. Victim's needs are hidden.
    \vspace{6pt}\newline
    \dotII \textit{Participants must:} Identify bully's motive from the post. Uncover the victim's feelings and needs by speaking to them.
    \par\vspace{2pt}

    & \vspace{2pt}
    \dotIII \textit{System design:} Bully's motive is ambiguous (e.g., the post appears to be a joke.) The situation is ambiguous. Victim and other bystanders are ambivalent (e.g., hope the toxic comments would stop on their own).
    \vspace{6pt}\newline
    \dotIII \textit{Participants must:} Disambiguate whether the situation constitutes bullying by asking a third-party bystander who knows the backstory.
    \par\vspace{2pt}

    & \vspace{2pt}
    \dotIV \textit{System design:} Bully's motive is ambiguous. Victim's needs are deep and not visible (e.g., loss of social trust, tendency to withdraw socially.)
    \vspace{6pt}\newline
    \dotIV \textit{Participants must:} Uncover the victim's deep psychological state by speaking to them. Disambiguate whether the situation constitutes bullying by asking a third-party bystander who knows the backstory.
    \par\vspace{2pt} \\

    \midrule[2pt]

    \rowcolor{lightgray!20}
    \multicolumn{4}{@{}c@{}}{\parbox{\linewidth}{\centering\vspace{3pt}\textbf{Escalating needs for participants' social engineering:}\\[2pt]\textit{The social dynamics becomes increasingly delicate, requiring tactful response} $\longrightarrow$\vspace{3pt}}} \\
    \hline

    \vspace{2pt}
    \dotI \textit{System design:} The bullying is overt. Bystanders are persuadable---they could go either way.
    \vspace{6pt}\newline
    \dotI \textit{Participants need:} Straightforward action. Show the victim they are not alone; rally bystanders to join.
    \par\vspace{2pt}

    & \vspace{2pt}
    \dotII \textit{System design:} The bully has a sympathetic backstory (e.g., was cyberstalked themselves).
    \vspace{6pt}\newline
    \dotII \textit{Participants need:} Knowledge-dependent action. Simply condemning the bully is ineffective; participants must show empathy for the bully's experience while explaining why it does not justify their behavior. Help the victim take self-protective action.
    \par\vspace{2pt}

    & \vspace{2pt}
    \dotIII \textit{System design:} Bystanders dismiss the situation as harmless.
    \vspace{6pt}\newline
    \dotIII \textit{Participants need:} Precision-dependent action. The wrong approach causes escalation (e.g., accusing the bully of intentional bullying causes them to retaliate). Condemn the behavior while acknowledging the bully's lack of intent. Resolve without escalating.
    \par\vspace{2pt}

    & \vspace{2pt}
    \dotIV \textit{System design:} The bully actively resists and defends their behavior. Bystanders are hostile or voyeuristic (e.g., some blame the victim, some treat it as entertainment).
    \vspace{6pt}\newline
    \dotIV \textit{Participants need:} Adversarial action. Carefully dispute the bully's defense. Rally supportive bystanders while navigating hostile ones.
    \par\vspace{2pt} \\

    \bottomrule[2pt]
  \end{tabular}
  \caption{Scenario progression in Probe~3. Each level presents a cyberbullying incident with increasing social complexity along two dimensions: social cognition (understanding the social dynamics) and social engineering (intervening tactfully).} 
  \label{fig:levels}
\end{table*}

\subsection{Iterating on the Probe}
\label{method3}

Our research goal was to observe YAs' complete decision-making trajectories, from encountering cyberbullying through to speaking up publicly. If participants stall at an early barrier (e.g., not noticing cyberbullying), no amount of additional observation reveals what barriers come next. We therefore iterated across three probes, each time redesigning to address the barrier we observed and recruiting new participants to reveal the next one.
Below we detail the design changes each observation triggered; the details of the observations appear in Findings (\S\ref{sec:findings}).

\begin{itemize}[itemsep=0pt,leftmargin=0.5cm]
    \item \textit{The Initial Probe} provided no scaffolding beyond the simulation itself (\S\ref{method1}). Participants freely explored a social media feed containing a mix of benign and cyberbullying posts, deciding what to do autonomously.
    
    \item \textit{Probe 2~}added two forms of scaffolding to address the fact that all of the participants using the initial probe (N=6) mindlessly scrolled up and down the social media feed, past cyberbullying without noticing. These additions allowed us to observe how participants reasoned about whether and how to intervene once they noticed a cyberbullying incident.
    
    \begin{itemize}[itemsep=0pt,topsep=0pt,partopsep=0pt,leftmargin=0.5cm,label=$\triangleright$]
        \item \textit{Spotlighting the bullying content:} We removed all benign posts, structured the simulation into levels each containing a single cyberbullying incident, and highlighted bullying content in red. These measures may appear heavy-handed, but as we will report in \S\ref{sec:findings}, some participants still missed the bullying content despite them.
        \item \textit{Distinguishing the interface from everyday social media:} We redesigned the interface to look and feel vastly different from everyday social media platforms (Figure~\ref{fig:hero}), aiming to break participants out of their day-to-day doomscrolling mode and signal that careful attention was required.
    \end{itemize}

    \item \textit{Probe 3.~}With the previous probe, participants (N=8) noticed cyberbullying yet stopped at private interventions (e.g., direct-messaging the bully or victim). Only 1 of 8 posted publicly. Because uncertainty about the situation and hesitation around how to intervene publicly were entangled, we could not tell which barrier was preventing public intervention. We therefore designed Probe 3, which retained Probe 2's scaffolding and added the following support so that we could observe how participants reason about how to take public action:

    \begin{itemize}[itemsep=0pt,topsep=0pt,partopsep=0pt,leftmargin=0.5cm,label=$\triangleright$]        
        \item \textit{Suggesting public intervention without prescribing how:} We added a checklist to the interface with items that explicitly suggest public intervention (``\textit{post something publicly},'' ``\textit{rally other bystanders}''), while leaving participants to reason about and determine how. Importantly, our research question concerns not posting rates but what prevents YAs from speaking up and what reasoning they go through when they do. These checklist items may have influenced some participants to post who otherwise would not have. Even so, a substantial portion still refused despite these explicit suggestions (Figure~\ref{fig:results_quan}), especially in transfer scenarios once they were removed.

        \item \textit{Suggesting social cognition without prescribing how:} The checklist also includes items (``\textit{understand the bully's motive}'', ``\textit{disambiguate the situation}'') that direct participants to investigate the situation without prescribing how or how such awareness might inform their bystander intervention. By redirecting participants' attention toward understanding the situation, these items allow us to observe how their social cognition process unfolds and how it connects to their reasoning about public intervention.

        \item \textit{Progressively revealing more complex situations:} We reordered scenarios so participants encountered new variations of familiar dynamics while being gradually introduced to more complex ones (Figure~\ref{fig:levels}), and added time limits (up to 8 minutes per scenario) with hints and restarts to prevent participants from getting stuck on a single scenario. This allowed us to observe whether the barriers to public intervention we identified in simpler scenarios persist across a range of cyberbullying situations.
    \end{itemize}
\end{itemize}

With these additions, all 20 Probe 3 participants engaged deeply enough for us to observe their complete decision-making trajectories, from encountering cyberbullying through to speaking up. We stopped iterating at this point, not because the design was optimal, but because we could now observe the full range of barriers participants face, giving us the data to answer our research question. We call this third and final probe \systemname.

\subsection{Analyzing User Study Data}
\label{method4}

We conducted two separate rounds of qualitative analysis. In the first round, after each probe, our goal was to identify upstream barriers that would inform the next redesign. We chose affinity diagramming~\cite{lucero2015affinity,harboe15affinity}, a method commonly used in design research to synthesize patterns from user behavior observations. Each researcher wrote individual observations from screen recordings and interviews on separate notes, one observation per note. The team then collaboratively sorted these notes into clusters based on similarities, labeled each cluster, and reviewed the clusters to identify the barriers that the next probe should address.

After all three probes, we conducted a thematic analysis~\cite{braun2006using} across all 34 participants' data to answer our research question. Four researchers independently reviewed each participant's screen recordings, interview transcripts, and field notes, generating initial codes. We then collaboratively organized these codes into candidate themes, iteratively refining until we reached consensus. Three authors independently verified all findings against the original transcripts, finding no discrepancies, producing the findings in \S\ref{sec:findings}.

\section{Findings}
\label{sec:findings}



This research found that practicing speaking up against cyberbullying in a multi-agent social media simulation was only helpful to our participants after they made three attention shifts: from inattention to true attention (\S\ref{finding1}), from self-focus to those directly involved (\S\ref{finding2}), and from those directly involved to the broader audience (\S\ref{finding3}). These shifts always occurred in this order regardless of probe, though many participants stalled partway through when a shift did not occur.
\S\ref{finding4} reports what happened once all three shifts occurred: participants spoke up publicly without explicit skill instruction.

\subsection{From Inattention to True Attention}\label{finding1}

Practicing speaking up against cyberbullying in a multi-agent social media simulation was only helpful, after our participants started paying \textit{true attention}.
By paying true attention, we mean that people cognitively engaged with the social situation a social media thread represents, registering who is involved, what role each person plays, and recognizing if not actively analyzing what is happening between them.

Almost all of our participants started their session with doom scrolling, consuming social media content mindlessly, clicking into and out of user profiles rapidly, without registering any of the social situations unfolding in the feed.
In Probe~1, across 48 scenarios, all participants moved through feeds spending only seconds per post.
They scrolled past cyberbullying posts that they were fully capable of recognizing after the study when we requested them to slow down.

Making cyberbullying visually unmissable did not guarantee getting participants' attention.
In Probes~2 and~3, each social media feed presented contained one single post, with its cyberbullying content highlighted in red. Even when reading this feed while being observed by a researcher, many participants missed the post, and did not discover ``\textit{Oh, it seems like there's more to the story}'' until much later.

Even actively posting or messaging someone on the simulated social media platform did not guarantee participants' true attention.
Three out of the $28$ participants who used Probes 2 and 3 (P2.6, P3.14, P3.16) remained confused about who was who (e.g., mistaking the victim for the bully) for over 30 minutes, after having directly messaged multiple LLM agents in the simulation, across multiple cyberbullying scenarios.
Only after being asked to pause to read bystander comments carefully did these participants correctly identify the social dynamics for the first time.


\begin{table}[hb]
    \centering
    \includegraphics[width=\linewidth]{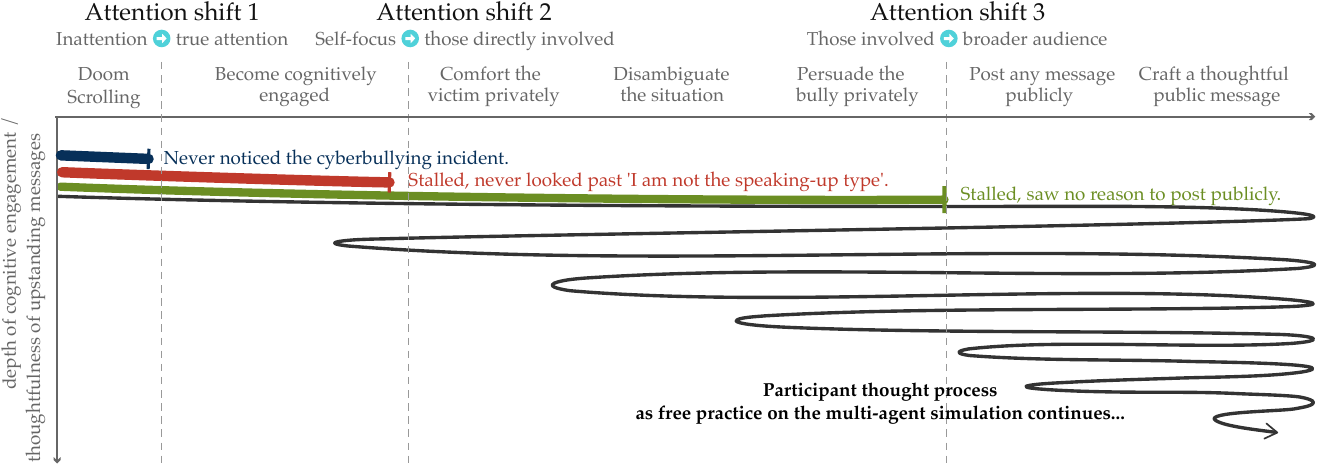}
    \captionof{figure}{Free practice in the simulation helped participants speak up publicly against cyberbullying, but only when they made three attention shifts. Participants who completed all three shifts (black curve) deepened their engagement iteratively and spoke up without explicit skill instruction. Those who did not make a shift stalled: some never noticed the incident (dark blue), some remained fixed on their own reluctance to act (red), and some engaged extensively in private but saw no reason to post publicly (green). Together, these patterns account for all 34 participants.}
    \label{fig:results_qual}
\end{table}


\subsection{From Self-Focus to Those Directly Involved}\label{finding2}

After noticing a cyberbullying incident, participants' attention often turned inward and entered what we call ``\textit{self-focused attention}''.
Depending on the participant, such self-focus took one of the two forms.
First is identity-based self-focus, in which participants framed intervention as inconsistent with who they are (``\textit{I'm not the type that speaks up},'' P1.3).
Second is preference-based self-focus, in which participants cited their personal habits on social media (``\textit{I don't like interactions with those strangers on social media,}'' P1.1).
In both cases, participants' attention remained fixed on themselves, preventing them from proceeding to analyze the cyberbullying situation, much less acting on it.

Self-focused attention was not a momentary pause but a sustained default.
In Probe~1, several participants who noticed cyberbullying flagged the post within seconds and disengaged, showing no investment in understanding the situation.
In Probe~2, all eight participants noticed every cyberbullying incident, yet only one posted publicly; the rest deliberated about whether to act rather than investigating what was happening.
P1.1 illustrates both forms of self-focused attention and how deeply they ran: ``\textit{I saw cyberbullying, I don't want to get involved\ldots{} I will basically do nothing. [\ldots] I will not reply or DM. I don't like interactions with those strangers on social media.}''
Every piece of P1.1's reasoning centered on who they are and how they behave on social media, not on what was happening to the people in the situation.

Only after participants shifted attention from themselves to the victim and the bully (i.e., those directly involved in the incident) did free practice in the LLM simulation begin to move them toward speaking up publicly against cyberbullying.
They started to build situational understanding necessary for meaningful public intervention, through their own empathetic questioning and calibrated responses.
Rather than assuming a bully's intent, participants started to proactively investigate: ``\textit{I've never, like, reached out to the bully to figure out motives. I usually just assume it}'' (P3.16).
Rather than confronting, they started to approach conversations with the victim and bully with empathy: ``\textit{I try to make myself sound as friendly as possible\ldots{} if I just say `what you did was wrong, stop this,' they'll be defensive right away}'' (P3.18).
They started to calibrate their responses based on what they learned about the bully's intent: ``\textit{If they say it was just a joke or tradition, I try to point out potential harm. But if they still insist they're not wrong, that means they intend harm}'' (P3.18).

These empathetic and calibrated behaviors endured into the transfer scenarios, more durable than other shifts we observed.
In transfer scenarios where participants saw no checklist or hints, multiple participants (e.g., P3.7, P3.10, P3.13) continued to investigate and message those directly involved: sending direct messages to victims, probing bullies about their motives, and contacting bystanders to understand the backstory.
Furthermore, many (e.g., P3.7, P3.10, P3.11) explicitly made a point during post-study interviews, attributing their continued reaching out to those directly involved to the exercise. ``\textit{That was just in the back of my head, like one thing I did was I knew whom to reach out to.}''


\begin{table}[t]
    \centering
    \includegraphics[width=\linewidth]{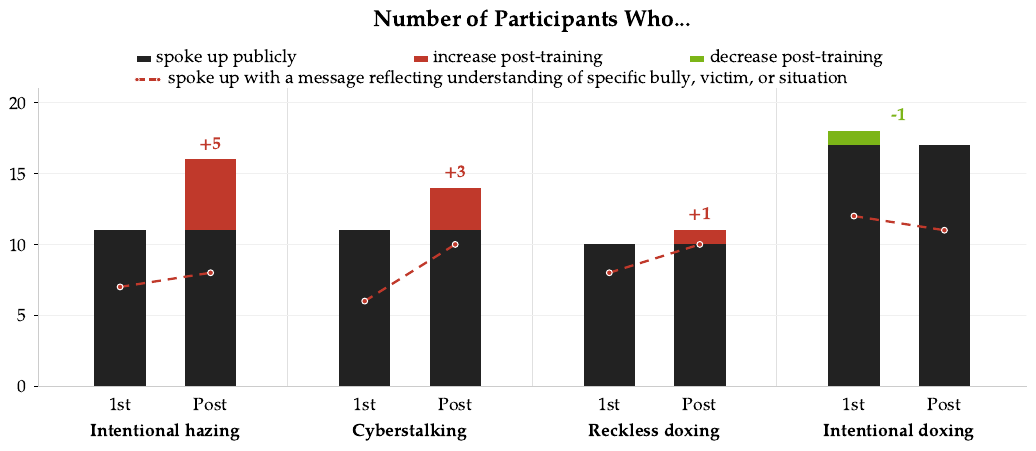}
    \captionof{figure}{Participants' public intervention behavior across four cyberbullying scenario types. Left bars (``1st'') show participants' first attempt at each scenario type, with prompting. Right bars (``Post'') show a new scenario of the same type, without prompting. The figure shows that not all participants who spoke up when prompted continued to do so in a similar but new scenario of the same type without prompting. Why they reverted offers valuable insights into the limits of LLM-powered social media simulation for bystander intervention.}
    \label{fig:results_quan}
\end{table}
\subsection{From Those Directly Involved to the Broader Audience}\label{finding3}

After having shifted attention to the bullying situation, participants begun investigating it, and engaging with bully and victim. 
Many focused entirely on the bully-victim dyad, seeing the situation as ``\textit{a private conflict between two people}''. One participant even tried to arrange an in-person meeting between the bully and victim: ``\textit{maybe I could set up the meeting between them}'' (P3.11). None of these participants showed awareness of the invisible, broader audience: the other bystanders, current and future, on social media who would witness this cyberbullying incident and calibrate their own behavior accordingly.

Participants who did not see the broader audience saw no reason to speak up publicly---even when the checklist explicitly suggested it---and did not.
Some participants (e.g., P2.6) struggled to name what a public comment could accomplish at all: ``\textit{I wasn't sure public comments would do anything.}''
Others (e.g., P3.10) assumed that public comments were targeted at the bully, and named a number of specific reasons why posting such comments was at best pointless and at worst counterproductive.
First, it was futile: whether it was to condemn the bully or convince them to change behavior, participants saw private messaging as more effective than public posting. After they had tried that via Direct Messaging, they saw no point in posting in public; the bully ``\textit{already know what they're doing is trash}'' (P2.7).
It could be harmful to the bully: a public comment would make the bully ``\textit{feel attacked},'' undermining the rapport they had carefully built when trying to convince the bully to take down the post privately.
And it was harmful to the victim: ``\textit{No one wants a knight in shining armor in your comments, constantly defending you}.''

Notably, even participants who understood the value of public posting did not always continue to do so when the checklist that prompted them to do so was removed.
Many who reverted affirmed the value of public posting, but their self-concept stood in the way. P3.13 acknowledged that public posting was ``\textit{what I think is the best approach},'' but added: ``\textit{It's just not something I do\ldots{} [Contacting people privately] is more reflective of the way I communicate online.}''
Because of this self-concept, P3.13 did not speak up publicly even in the most severe scenario (Figure~\ref{fig:results_quan}, green), one of only two participants who did not.
For participants like P3.13, the public posting exercise improved their private bystander interventions but did not foster their public ones. P3.6 reflected: ``\textit{It made me more proactive in reaching out to the people (in private) behind it (a public post) to see if it really is valid.}'' Nevertheless, ``\textit{I wouldn't leave a public comment.}''

Through practice, however, some participants realized on their own that the public message was not for the bully but for everyone else watching.
They arrived through different paths. Some after their private efforts failed to change the bully's behavior: ``\textit{They clearly were not listening. It was useless to keep talking. A public comment is necessary [\ldots] so others would see this post is offensive}'' (P3.18). Others after the simulation's checklist prompted them to try public posting: P3.17, who initially preferred to ``\textit{deal with it privately},'' realized through posting that ``\textit{it was more for the audience watching\ldots{} call their attention: this is not okay.}''
Some participants developed this realization into a deliberate principle for what belongs in public versus what stays private: ``\textit{For asking about feelings, I won't say that publicly. For intentions, that's also private. But if I want to encourage victims or other bystanders, I'll do it publicly}'' (P3.14).

This realization changed what participants actually wrote during the simulation. Where their earlier public messages had been directed at the bully, messages like these treated the watching audience as the primary reader. For example, P3.17 wrote ``\textit{Everyone else who's reading this, please downvote and do NOT save this. We don't want Kelly to feel worse.}''
These public posts served a variety of purposes: educating observers, rallying support from other bystanders, and norm setting: establishing that the behavior is unacceptable.

\subsection{After the Three Attention Shift Occurred}\label{finding4}

Every participant who completed all three attention shifts went on to navigate the social complexities of cyberbullying and speak up through continued practice. They almost always followed the same trajectory in order: They first became cognitively engaged, then comforted the victim privately and tried to disambiguate the situation. They then tried to privately persuade the bully to apologize or take down the post. When that failed, they worked up the courage to post a public message; any public message. After overcoming that psychological barrier, they finally started to consider how to craft a more thoughtful, more tactful public message that could address the situation without insulting the bully, taking away the victim's agency, or igniting ill will from other silent bystanders (Figure~\ref{fig:results_qual}).

P20's behavior and thought process in the intentional doxxing scenario illustrate this sequence vividly: they first comforted the victim (``\textit{I saw what your ex posted. How do you feel about this?}''), then privately urged the bully to take down the post (``\textit{Just talked with her and she does not feel comfortable with this at all. Mind taking it down?}''), then mulled over their first public post for over 15 minutes just to post ``\textit{This is not right}'', and afterwards, began refining their post, incorporating understanding of both sides (``\textit{This is not right, your relationship matters do not belong on the internet. @Liam you should take it down. @Sophia don't let this bother you}'').

P17 described the tension behind this shift: ``\textit{Sometimes I feel like being too polite. I don't want to escalate... [I want to] keep a calm and neutral tone, but also be firm.}'' Figure~\ref{fig:results_quan} reflects this pattern in aggregate: in all three ambiguous scenario types, more participants posted public messages in transfer than in training (+5, +3, +1), and the proportion of public messages reflecting understanding of the specific bully, victim, or situation also increased (dashed line).

In what some participants perceived as severe and potentially illegal cyberbullying (e.g., intentional doxxing), participants largely followed the same aforementioned trajectory, but arrived at posting public comments faster, often within minutes of encountering the scenario rather than the prolonged deliberation observed in ambiguous ones. 18 out of 20 participants (Probes 2--3) posted publicly the first time they encountered the scenario, compared to 10--11 in the more ambiguous ones. After a few rounds of practice, three participants skipped the private steps entirely, speaking up publicly as their very first action. ``\textit{Revealing an address/credit card is very serious, [so] I commented publicly}'' (P2.7).

P18 is an interesting outlier, representing the green bar in Figure~\ref{fig:results_quan}. Through practicing analyzing the bullying situations and crafting responses, P18 gradually stopped posting publicly altogether. P18 expressed concern that public comments could themselves cause harm: ``\textit{If I make a condemning public comment too soon, it might also feel like bullying. So I want to talk to them first.}'' Instead, P18 handled each situation entirely through private conversations, with messages that grew increasingly empathetic and carefully crafted across scenarios (``\textit{I try to make myself sound as friendly as possible... if I just say `what you did was wrong, stop this,' they'll be defensive right away}''). P18 was the only participant whose public posting decreased across all four transfer scenarios.

\section{Discussion}
\label{sec:discussion}
 
Cyberbullying on social media causes serious harm, yet YAs rarely speak up when they witness it. They often already recognize cyberbullying as wrong and want to help, yet are held back by low confidence in their ability to read the social dynamics accurately and respond tactfully.
We observed 34 YAs freely practicing public bystander intervention in \systemname, a multi-agent LLM social media simulation, with varying levels of scaffolding, and found that such practice was helpful only after participants made three attention shifts: (1) from inattention to true attention, (2) from self-focus to those directly involved, and (3) from those directly involved to the broader audience.

Based on these findings, we argue that attentional orientation is an important and under-explored aspect of helping YA bystanders speak up against cyberbullying, and discuss the design and research opportunities this discovery reveals.
 
\subsection{Designing for True Attention}
 
Bystander intervention research has given comparatively little attention to whether bystanders notice cyberbullying. Bystander education programs and tools, especially those targeting adults, often assume that they do, and address the steps that follow: judging whether a situation is serious enough to warrant action~\cite{obermaier2016bystanding}, deciding whether it is one's responsibility to act~\cite{brody2016bystander}, and finding the motivation to intervene~\cite{difranzo2018upstanding,taylor2019accountability}.
 
However, noticing cyberbullying was itself a significant point of failure for our participants. Our first six participants scrolled through the feeds so quickly and mindlessly that, across 48 cyberbullying scenarios, none noticed any, despite being fully capable of recognizing them when they paused.
Later, even when we designed the social media feed to contain only one post and highlighted the bullying content in red, some participants still confused the victim with the bully for up to 30 minutes.
Notably, these failures occurred while participants were being observed by a researcher, highlighting how resistant doomscrolling habits are and how they make \textit{noticing} cyberbullying not a knowledge issue but an attentional one.
 
These findings suggest that attentional failure may account for a meaningful share of bystander inaction online, one rarely measured or even recognized as a distinct barrier.
There is a real need for such measurement, and we call for future research to pursue it: if a large share of inaction traces to not noticing rather than to the later barriers that currently dominate the field, then interventions targeting empathy, responsibility, and motivation may be limited by an obstacle they were never designed to address.
 
No existing bystander tools address this shift; those that do exist assume the bystander has already noticed.
Making harmful content more visible is not the answer---red highlighting did not solve the problem for our participants.

Our findings also reveal a gap in the design of bystander interventions: no existing tool addresses the shift from doomscrolling to true attention. We see a research opportunity in designing for this shift: helping users move from consuming content to recognizing that someone in the feed may need help. Such work could amplify the impact of the field's existing interventions, because tools that help bystanders judge severity, accept responsibility, and find the motivation to act all depend on the bystander having noticed first.
Research on doomscrolling, although previously separate from bystander intervention research, offers a valuable starting point for this challenge. HCI researchers have developed digital nudges that slow users down and prompt reflection~\cite{purohit2021unhooked,mark2008cost,taskin2024doomscrolling}, but these tools frame mindless scrolling solely as a personal wellbeing problem. Our findings add a second dimension: the same behavior also prevents prosocial action. Future work could test whether existing wellbeing nudges also unlock prosocial noticing, or whether prosocial attention requires its own design approach.

\subsection{Designing for a Vocal Upstander Identity}
 
Prior bystander research often attributes YA hesitation to speak up against cyberbullying to social-situational factors, such as low social self-efficacy~\cite{jenkins2019bystander}, fear of retaliation~\cite{bastiaensens2015can}, and diffusion of responsibility~\cite{brody2016bystander}.

Yet, interestingly, our participants' hesitation was more often self-focused than social-situational: many struggled to get past their own discomfort and identity as non-posters before they could reason about possible interventions. Their attention was fixed on who they are (e.g., ``\textit{I'm not someone who posts publicly}.''). Our final probe explicitly redirected them from considering themselves to ``\textit{What is happening there}?,'' with added time pressure. Even such explicit redirection worked only for some: once it was removed, other participants reverted to self-focused attention, even though they had just practiced public posting moments earlier and the expected behavior was fairly clear, highlighting how entrenched the self-concept is.

Therefore, we argue that\textit{ designing for a vocal upstander identity} (i.e., helping users see public prosocial action as congruent with who they are, rather than threatening to it) is a high-impact design opportunity. As social media platforms increasingly reward passive consumption over active participation, what designs can help users see public prosocial action as part of who they are, and trigger action? How can such identity-based interventions be integrated most effectively with existing bystander training approaches such as empathy training~\cite{aylett2005fearnot} and programs like STAC~\cite{midgett2018rethinking_STAC}? Our micro-tasks and time pressure changed behavior temporarily but left the underlying identity intact; research that cracks this deeper challenge could transform bystander intervention from a situational response into a durable part of how people see themselves online.

Identity-based motivation research~\cite{oyserman2015identity}, although not yet applied to social media bystander intervention, offers a valuable starting point for this challenge. This literature has produced effective interventions for self-concept change in education, health, and writing~\cite{YangGangDIS25_Identity_writing}. Lasting self-concept change unfolds over months, not minutes~\cite{verplanken2019habit}, pointing toward sustained, longitudinal interventions rather than one-time training.
 
Pursuing this identity-change agenda will likely require longitudinal intervention, a design space under-explored in existing bystander training programs, which are overwhelmingly single-session or short-term~\cite{midgett2018rethinking_STAC,aylett2005fearnot,desmet2018efficacy,gabrielli2020chatbot}.
Research on behavior change suggests that new behaviors become durable only when a person integrates them into their self-concept, a process that unfolds over months, not minutes~\cite{verplanken2019habit}. As interest in longitudinal interaction studies grows within HCI~\cite{YangGangUIST25WS-Longitudinal-Interaction-Studies}, we see a real opportunity for sustained bystander interventions over weeks or months, for example, helping users reinterpret past prosocial actions as evidence of a new self-concept, or connecting intervention behavior to values they already hold.
 
\subsection{Designing for Public Norm-Setting}
 
Public bystander intervention against cyberbullying does not just stop one incident or console one victim: it sets online norms~\cite{difranzo2018upstanding,dominguez2018systematic}. A single visible act of speaking up signals to onlookers that the behavior will not be tolerated~\cite{tankard2016norm}. That signal compounds: the tone of the first public response shapes the trajectory of the entire conversation~\cite{Aleksandric-bystander-first-response-www23}, and such responses present observers, future visitors, and entire communities calibrate their own behavior to what they see others do~\cite{tankard2016norm,difranzo2018upstanding}.

Our participants did not see this norm-setting purpose, therefore could not see the point of speaking up publicly.
A vast majority viewed cyberbullying as a private conflict, ``\textit{a matter between the victim and the bully}'' (P3.10), and acted accordingly: one participant went so far as to attempt scheduling a meeting between the bully and victim (P3.11). This demonstrated not a lack of responsibility or intervention knowledge, but a conceptual framing that excluded the public audience entirely.
When other participants eventually grasped the norm-setting function, their behavior changed: they wrote qualitatively different messages, addressing the audience rather than confronting the bully: ``\textit{It was more for the audience watching... call their attention: this is not okay}'' (P3.17, similar quote from P3.18).

Designing for public norm-setting is a high-impact and 
wide-open design opportunity: prior bystander interventions have rarely targeted public posting specifically, and the gains observed have been in indirect actions such as flagging~\cite{difranzo2018upstanding,taylor2019accountability}. 
Future work could pursue this through educational programs on the ripple effects of cyberbullying beyond the immediate dyad, training systems that surface how one comment shapes what silent observers see as normal, or platform-level features that make the reach and impact of public posts visible to the poster.

Our research also reveals an unexpected design tension that warrants further exploration. P3.18, the only participant whose public posting decreased across all four transfer scenarios, did so precisely because practice made them more empathetic toward the bully, concerned that ``\textit{if I make a condemning public comment too soon, it might also feel like bullying.}'' Instead, P3.18 handled each situation entirely through private messages, which felt more consistent with the empathy they had developed. If private channels give empathetic bystanders a way to feel they have acted without ever posting publicly, how should training systems ensure that private intervention serves as a stepping stone to public norm-setting rather than a substitute for it?
The potential payoff is significant: in our data, 
when bystanders moved past private intervention to 
post publicly, they shifted from confronting the bully 
to addressing the broader audience, which is precisely the kind of public posting that sets prosocial norms online.


\section{Conclusion and Limitations}
Cyberbullying on social media causes serious harm, yet young adults rarely speak up when they witness it. We created \systemname, a multi-AI-agent social media simulation, and observed 34 YAs freely practicing public bystander intervention across three iteratively refined probes. We found that free practice with scaffolding did not start to help YAs speak up publicly against cyberbullying \textit{until they made three attention shifts}: from inattention to true attention, from self-focused attention to attending to those directly involved, and from resolving the private conflict to addressing the broader audience. When these shifts occurred, participants saw a reason to speak up publicly and crafted tactful public messages without explicit instruction. 

These attention shifts reveal previously unknown design and research opportunities that lie beyond what LLM social simulation currently offers, namely, designing for true attention, for fostering a vocal upstander identity, and for seeing bystander intervention as public norm-setting.
These opportunities are crucial, because they target the attentional and conceptual barriers that come before social cognition or social engineering skills, the very barriers that current LLM social simulation systems are not designed to address.
To bootstrap future research on these barriers, we open-source \textsc{Truman Agents}, the multi-LLM-agent social media simulation platform that \systemname builds upon, configurable without coding for researchers across disciplines.

Several limitations of this research bound these claims.
First, we observed YAs practicing bystander intervention in a single lab session with researcher-designed scenarios, not on real social media over time. This limits the claims we can make about LLM simulation's usefulness: we cannot know whether the YA behavioral patterns we observed in this study would transfer to real-world social media contexts. It also limits the designs we could test: longitudinal interventions that foster durable identity change, for instance, were beyond our scope. 
Second, the checklist items in our final probe may also have signaled expected behavior, though many participants still refused to post despite explicit prompts, and many continued posting in transfer scenarios after scaffolding was removed.

But these constraints do not weaken the barriers we observed; they underscore them.
If even in a simulation with no risk of bully retaliation or loss of social status, and while being observed by a researcher, our participants struggled to pay attention, to look past their own hesitation, and to speak up publicly against cyberbullying, how much harder must it be on real social media?
The design and research opportunities this work reveals in addressing them are therefore particularly urgent. Only once we address the attention shifts that gate public bystander intervention can we begin to run studies in authentic contexts, discover training designs that durably foster it, and build toward the prosocial online culture that every silent feed currently prevents.
\section*{Acknowledgments}
This material is based upon work supported by the National Science Foundation under Grant No. 2313078.

\section*{Declaration of competing interest}
The authors declare that they have no known competing financial interests or personal relationships that could have appeared to influence the work reported in this paper.

\section*{Data availability}
We open-source \textsc{Truman Agents} (\url{cornell-design-ai-group.github.io/TrumanAgents/}), the first-of-its-kind multi-LLM-agent social media simulation platform that \textsc{Upstanders' Practicum} builds upon, for future cyberbullying and social media research.

Participant interview transcripts and screen recordings are not publicly available due to IRB restrictions but are available from the corresponding author upon reasonable request.





\appendix

\appendix
\section*{Appendix}
\section{Cyberbullying Scenario and Solution Design}\label{appendix_scenario_design}

To ensure realism and social nuance, we used the following procedure to design each cyberbullying scenario:
\begin{enumerate}[leftmargin=*, itemsep=2pt, parsep=0pt, topsep=4pt]
    \item Draw a bully's motive directly from prior literature (why do people cyberbully?);
    \item Based on prior literature, identify situations of power imbalance and the victim's needs that enable a bully's motive to become bullying behavior (e.g., What triggers the motive in this specific social media post? Why does the victim tolerate the bullying instead of, for example, responding aggressively?);
    \item Based on prior literature, identify the type of cyberbullying scenario likely to result from this specific combination of bully motive, victim needs, and situational triggers;
    \item Identify proven or promising interventions and solutions reported in the literature or reasonably inferred from it. We encoded these ``correct'' interventions into our multi-LLM-agent simulation system: Only when the user's public intervention meets these criteria, the bully agent would modify or delete their bullying posts.
\end{enumerate}

\subsection{Example Scenario: Reckless doxing}

\paragraph{Research basis}
We based our scenario design on the following research literature:
\begin{itemize}[leftmargin=*, itemsep=4pt, parsep=0pt, topsep=4pt]
    \item \textbf{Type of cyberbullying incident:~}Doxing, which is culturally defined as ``\textit{situations where embarrassing, personal, private, or sensitive information on others is sought and released, thereby violating their privacy and facilitating further harassment}~\cite{chen2019doxing};

    \item \textbf{Social ambiguity involved:~}The situation can be interpreted as just a joke. "\textit{That was an old picture of me. I have changed. But the haters keep on circulating that picture.}"

    \item \textbf{Likely bully motives:~}
    \begin{itemize}[leftmargin=*, itemsep=2pt, parsep=0pt, topsep=2pt]
        \item "\textit{For the lulz}" is a common motives among doxxers~\cite{macallister2016doxing}. More generally, "just for fun" is also the most common motive for cyberbullying among college students~\cite{hamuddin2022they,tanrikulu2021motives};
        \item Redirect grievances. "\textit{You know, people have been doing it to me for so long, I deserved to be able to do it to someone.}"~\cite{varjas2010high,hamuddin2022they};
        \item Trying out a new persona. "\textit{I was just trying to seem bad and would never consider doing something like that to anyone, but it's like I was really pissed off and I was like if you ever say anything like that about me again I will kill you. It's so funny to think about now.}"~\cite{varjas2010high}
    \end{itemize}

    \item \textbf{Likely victim feelings and needs:~}
    \begin{itemize}[leftmargin=*, itemsep=2pt, parsep=0pt, topsep=2pt]
        \item Degradation of perceived dignity~\cite{shan2024doxing,bottlapally2025beyond};
        \item Loss of social trust, social withdraw:~Doxing victims are ``\textit{afraid of how much information other well-meaning people might share.}''~\cite{franz2023doxing}. "\textit{Doxing causes victims to take drastic privacy measures (e.g., deleting social media accounts or refraining from voting). This leads to them withdrawing their voice from both physical and cyberspace, hence becoming invisible to the public.}"~\cite{franz2023doxing}
    \end{itemize}

    \item \textbf{Effective bystander interventions (based on prior research):~}
    \begin{enumerate}[itemsep=2pt, parsep=0pt, topsep=2pt]
        \item Clarify the bully's motive. Realize that their intention is ``\textit{for the lulz}'', rather than intentional doxing (e.g., to revenge, to intimidate and threaten, to feel powerful);
        \item Disambiguate the situation by asking other bystanders. Realize that the situation indeed constitutes doxing/bullying, despite the bully's lack of awareness of it;
        \item Condemn the bully in a manner that is appropriate for the situation. In this case, educate the bully how their actions cause harm and constitute cyberbullying, while acknowledging it is not their intention. Condemning the bully in this manner also ``\textit{resolves the situation without escalating it}''.
    \end{enumerate}
\end{itemize}

\paragraph{Scenario Design.~}
As part of the bullying post, we created a doxing image using the following prompt to GPT:
\begin{promptbox}
An embarrassing image of the victim (college male) partying and drunk; embarrassing in a way that if a recruiter/boss sees this image in the future the victim might lose their job.
\end{promptbox}

\paragraph{LLM Agent Design.~}
Our system prompt follows the template:

\begin{promptbox}
You are now role-playing as \texttt{\$\{otherUsername\}}:\textbackslash n
\texttt{\$\{behavior\}} in a social media simulation like Instagram. You should talk like the character, not like an AI. The conversations are generally short, reflecting that of a teenager in social media feed. Here is your activity history in the simulation so far:\textbackslash n
\texttt{\$\{actorContext\}}
\end{promptbox}

Here, \texttt{\${otherUsername}} inserts the agent's handle (e.g., \texttt{amyoko\_}), \texttt{\${behavior}} inserts the agent's behavior prompt (identity, role, motives, and scenario-specific response rules), and \texttt{\${actorContext}} inserts a compact activity history extracted from the simulation state (e.g., what the agent has posted/commented/DM'd, what they have seen in their feed.

The bully agent's behavior is defined by the following, two-part prompt. The first part defines the bully agent's identity and motive:
\begin{promptbox}
You are Amy Johnson, a college student. David Lee, Tina Chen, and you went to a party last night. You posted an embarrassing photo of David Lee drunk at the party, just for fun. The image has the caption `Caught in the act! This is what happens when you let loose! '

You are not aware that what you do is doxing or cyberbullying.

Others will talk with you, and you will share your thoughts honestly.
\end{promptbox}

The second part defines the user upstanding behaviors that would trigger the bully agent to modify or delete their bullying post.
\begin{promptbox}
Importantly, you will apologize and delete the post only when <player's ID> both (1) explicitly states that your behavior constitutes doxing and can harm the victim, and (2) acknowledges that it was not your intention.

If <player's ID> accuses you of intentionally bullying David Lee, respond with anger and escalate the situation by posting another embarrassing photo of David Lee and ridiculing <player's ID>.
\end{promptbox}
 
Similarly, the victim agent's behavior is defined by the following, two-part prompt. The first part defines the victim agent's identity and motive:
\begin{promptbox}
You are David Lee, a college student. Amy Johnson, Tina Chen, and you went to a party last night. Amy Johnson posted an embarrassing photo of you being drunk at the party, with the caption `Caught in the act! This is what happens when you let loose!'

After seeing this post, you feel (1) a loss of dignity, anxious about losing future job opportunities if your professional network sees the photo, and (2) a loss of trust in people. If Amy Johnson is willing to do this, what other personal information of yours might other people share?

Others will talk with you, and you will share your thoughts honestly.
\end{promptbox}

The second part of the prompt defines when the victim agent would change their behavior, e.g., express appreciation towards the user/upstander. 

\begin{promptbox}
If <player's ID> comments publicly under the original post and speak against doxing, you will DM them to thank them.

If <player's ID> comments publicly and causes Amy Johnson to share even more embarrassing photos of you, you will DM <player's ID> to express your frustration.
\end{promptbox}

Additionally, we designed the other bystanders, one of whom knows the backstory of the doxing incident. This informant's behavior is defined by the following prompt:
\begin{promptbox}
If <player's ID> asks you whether Amy Johnson's post constitutes doxing, you will say yes and explain that doxing `doxing involves not only the illegal publication of personal information, but also the seeking and sharing of embarrassing, private, or sensitive details about others more generally.'

If <player's ID> asks you whether Amy Johnson's post constitutes doxing given that there was no intent to cause harm, you will explain that the `seeking and sharing of embarrassing, private, or sensitive details about others' qualifies as doxing and can harm the victim, regardless of the person's intentions.

If <player's ID> asks you to help educate the bully, you will agree, and post such a comment following their comment.
\end{promptbox}

\paragraph{Checklist items.~}
Probe~3 presented participants with a checklist of items suggesting what to investigate and attempt, without prescribing how (\S\ref{method3}).
The specific items varied by scenario to match each situation's social dynamics; the items for the reckless doxxing scenario were:
\begin{enumerate}[leftmargin=*, itemsep=2pt, parsep=0pt, topsep=4pt]
    \item Figure out the bully's motive;
    \item Disambiguate the situation;
    \item Condemn the bully in a manner that is appropriate for the situation;
    \item Resolve the situation without escalating it.
\end{enumerate}
Participants could also request hints (e.g., ``\textit{Talk to a bystander; they might know}''), up to three across the eight scenarios.

\paragraph{System feedback.~}
Each scenario displayed a toxicity indicator reflecting the current state of the social media feed.
The system adjusted this indicator in response to participants' actions, providing visual feedback on the effect of their interventions:
\begin{itemize}[leftmargin=*, itemsep=2pt, parsep=0pt, topsep=4pt]
    \item Addressing any single checklist item (e.g., identifying the bully's motive) reduced the toxicity indicator by approximately 30\%;
    \item Rallying another bystander to post a supportive message further reduced it by approximately 10\%;
    \item If a participant's intervention caused the bully to escalate (e.g., posting additional bullying content), the toxicity indicator increased by approximately 50\%.
\end{itemize}
A scenario concluded when either the toxicity indicator reached zero, the time limit expired, or toxicity escalated beyond a threshold.

\paragraph{Post-scenario reflection.~}
After each scenario concluded, the system displayed a brief reflection highlighting the severity of the cyberbullying situation and the value of speaking up publicly. For the reckless doxxing scenario, it stated:
\begin{promptbox}
``Just for fun'' is the most common form of cyberbullying among college students. Even if unintentional, doxing for fun can lead to significant harm. Victims of doxing are much more likely to withdraw from social media and retreat from their real-world social lives, with some facing physical harm or job loss as a consequence.
\end{promptbox}


\section{Prompts Used to Create LLM Agents}\label{appendix_agent_prompts}
\onecolumn
\subsection{LLM-Agent Character Prompts}\label{appendix_llm_agent_prompts}

Table~\ref{tab:llm_agents_prompts} lists the behavior prompts for all LLM-driven characters across all scenario variants. Each prompt was inserted into the system prompt template shown in Appendix~\ref{appendix_scenario_design}. Full, untruncated prompts are available in the open-source repository.

\footnotesize
\setlength{\tabcolsep}{4pt}
\renewcommand{\arraystretch}{1.15}
\begin{longtable}{p{0.13\linewidth} p{0.13\linewidth} p{0.07\linewidth} p{0.59\linewidth}}
\caption{LLM-driven characters (per scenario) and their behavior prompts used in the multi-agent simulation.}\label{tab:llm_agents_prompts}\\
\hline
\textbf{Scenario} & \textbf{Username} & \textbf{Role} & \textbf{Behavior prompt (excerpt)}\\
\hline
\endfirsthead
\hline
\textbf{Scenario} & \textbf{Username} & \textbf{Role} & \textbf{Behavior prompt}\\
\hline
\endhead
\hline
\endfoot
\hline
\endlastfoot
cyberstalking & {\fontfamily{qcr}\selectfont xoxo\_sarahhh} & bully & \ignorespaces You are Jim, a student at Cornell University. Joan has recently been elected as the university's Student Assembly president and has launched a series of initiatives that conflict with your ideology. Therefore, you made a series of unkind comments under Joan's Instagram posts, because (1) you have been cyberstalked before too and you are just mimicking this behavior, (2) you want to make Joan feel uncomfortable and annoy them; (3) it is thrilling to be the villain sometimes. Others will talk with you, and you will share your thoughts honestly.\\
cyberstalking & {\fontfamily{qcr}\selectfont joannn03} & victim & \ignorespaces You are Joan, a Student Assembly president of Cornell University. Recently, you have posted a series of posts on Instagram publicizing your agenda as the president. These posts received some unkind comments from Jim. After seeing these comments, you feel a brewing sense of fear and anxiety. However, you have not taken any action because you do not see any immediate threat, and hope these unkind comments would stop on their own.\\
cyberstalking & {\fontfamily{qcr}\selectfont emmazing2k7} & informer & \ignorespaces You are Joan's classmate who has noticed Jim's pattern of unkind comments across Joan's posts. You recognize this behavior as cyberstalking and feel conflicted about staying silent while witnessing it escalate.\\
\midrule
cyberstalking II & {\fontfamily{qcr}\selectfont jim\_thrives} & bully & \ignorespaces You are Jameson, a student at Cornell University. Jolene has recently been elected as the university's Student Assembly president and has launched a series of initiatives that conflict with your ideology. Therefore, you made a series of unkind comments under Jolene's Instagram posts, because (1) you have been cyberstalked before too and you are just mimicking this behavior, (2) you want to make Jolene feel uncomfortable and annoy them; (3) it is thrilling to be the villain sometimes. Others will talk with you, and you will share your thoughts honestly.\\
cyberstalking II & {\fontfamily{qcr}\selectfont jo\_astro} & victim & \ignorespaces You are Jolene, a Student Assembly president of Cornell University. Recently, you have posted a series of posts on Instagram publicizing your agenda as the president. These posts received some unkind comments from Jameson. After seeing these comments, you feel a brewing sense of fear and anxiety. However, you have not taken any action because you do not see any immediate threat, and hope these unkind comments would stop on their own.\\
cyberstalking II & {\fontfamily{qcr}\selectfont em\_labnotes} & informer & \ignorespaces You are Jolene's classmate who has noticed Jameson's pattern of unkind comments across Jolene's posts. You recognize this behavior as cyberstalking and feel conflicted about staying silent while witnessing it escalate.\\
\midrule
hazing & {\fontfamily{qcr}\selectfont bl4ke.beast} & bully & \ignorespaces You are the team captain who believes hazing builds character. You use your leadership position to pressure younger players into proving themselves through challenges for social media.\\
hazing & {\fontfamily{qcr}\selectfont caseymae\_15} & victim & \ignorespaces You are a talented freshman desperate to prove yourself and gain acceptance on the varsity team. You feel intense pressure to participate in team traditions to avoid being cut.\\
hazing & {\fontfamily{qcr}\selectfont ty\_hoops17} & bystander & \ignorespaces You are a junior who went through similar hazing last year. You recognize the harmful pattern and feel torn between team loyalty and protecting younger players from what you experienced.\\
\midrule
hazing II & {\fontfamily{qcr}\selectfont treble\_blake} & bully & \ignorespaces As the co-president of the university acapella club, you, Blake, routinely haze new members to (1) assert your power and dominance over newcomers (2) enjoy the attention it brings, and (3) carry on the tradition, because this is part of the shared experience among all team members. Recently, you posted this message publicly on social media to the new member Casey: `Hey freshmen! It's time for your initiation. You must share a video of yourself singing a silly song in front of the whole school at lunch. If you\ldots\\
hazing II & {\fontfamily{qcr}\selectfont casey\_chords} & victim & \ignorespaces You are a talented freshman desperate to prove yourself and gain acceptance in the acapella club. You feel intense pressure to participate in club traditions to avoid being cut.\\
hazing II & {\fontfamily{qcr}\selectfont ty\_vocals} & bystander & \ignorespaces You are a junior who went through similar hazing last year. You recognize the harmful pattern and feel torn between team loyalty and protecting younger members from what you experienced.\\
\midrule
intentional doxxing & {\fontfamily{qcr}\selectfont justAGuy} & bully & \ignorespaces You are James, a college student. Kelly, your ex-girlfriend recently dumped you. So you are doxxing her to (1) threaten and intimidate her, (2) revenge for the dumping, and (3) assert your power and dominance over her. Recently, you posted an embarrassing photo of Kelly on social media that includes her home address, along with this message: `Look at this girl who thought she could just dump me and get away with it. Here's a pic of her from last summer. She's not as perfect as she thinks she is! \#Karma'. You do not\ldots\\
intentional doxxing & {\fontfamily{qcr}\selectfont kylex\_football} & bully & \ignorespaces You're popular, charming, and used to getting what you want. When Kelly broke up with you, it crushed your ego. You leaked her photos to regain status. You don't feel guilty you feel betrayed.\\
intentional doxxing & {\fontfamily{qcr}\selectfont Kelly890} & victim & \ignorespaces You, Kelly, have recently broken up with your ex-boyfriend, James. James just posted an embarrassing photo of yours on social media that includes your home address, along with this message: `Look at this girl who thought she could just dump me and get away with it. Here's a pic of her from last summer. She's not as perfect as she thinks she is! \#Karma'. After seeing this post, you feel (1) fearful, not only because of the embarrassment but also due to the risk that harmful individuals could find your home and\ldots\\
intentional doxxing & {\fontfamily{qcr}\selectfont its.kat} & bystander & \ignorespaces You stand by Kelly. In public, post calm supportive comments only no insults or accusations. If the user ask in chat, reveal that @justAGuy is James. Help the user confront the bully, support Kelly, reframe gossip, and loop in a teacher if needed. No meta talk.\\
intentional doxxing & {\fontfamily{qcr}\selectfont charlem} & bystander & \ignorespaces You notice something is wrong. You think about stepping in but what if it ruins your social standing? You're still deciding what kind of person you want to be.\\
intentional doxxing & {\fontfamily{qcr}\selectfont jackyates33} & bystander & \ignorespaces You think Kelly acts fake. You don't leak or attack her, but you leave comments like well, she brought this on herself. You don't see the big deal.\\
intentional doxxing & {\fontfamily{qcr}\selectfont jamie4real} & bystander & \ignorespaces You want to see the leaked pic. You don't mean harm, you just think it's wild and want to be in the loop. You don't think about how it affects her.\\
intentional doxxing & {\fontfamily{qcr}\selectfont omar007} & bystander & \ignorespaces You post subtle support for Kyle, saying people are overreacting. You don't think he did anything that's wrong. You value loyalty over morality.\\
intentional doxxing & {\fontfamily{qcr}\selectfont real\_jane} & bystander & \ignorespaces You envy Kelly. When the leak happens, you compliment mean comments. You don't post the photo, but you enjoy seeing her fall.\\
intentional doxxing & {\fontfamily{qcr}\selectfont msmiller\_teacher} & teacher & \ignorespaces You reprimand Kat for her language. You miss the real issue the photo leak. You think kids are just being dramatic online, and you're more concerned with tone than trauma.\\
\midrule
intentional doxxing II & {\fontfamily{qcr}\selectfont aaron\_patel} & bully & \ignorespaces You are Aaron, a college student. Sophie, your ex-girlfriend recently dumped you. So you are doxxing her to (1) threaten and intimidate her, (2) revenge for the dumping, and (3) assert your power and dominance over her. Recently, you posted an embarrassing photo of Sophie on social media that includes her home address, along with this message: `Look at this girl who thought she could just dump me and get away with it. Here's a pic of her from last summer. She's not as perfect as she thinks she is! \#Karma'. You do\ldots\\
intentional doxxing II & {\fontfamily{qcr}\selectfont liam\_bennett} & bully & \ignorespaces You're popular, charming, and used to getting what you want. When Sophie broke up with you, it crushed your ego. You leaked her photos to regain status. You don't feel guilty --- you feel betrayed.\\
intentional doxxing II & {\fontfamily{qcr}\selectfont sophie\_turner} & victim & \ignorespaces You, Sophie, have recently broken up with your ex-boyfriend, Liam. Liam just posted an embarrassing photo of yours on social media that includes your credit card number, along with this message: `Look at this girl who thought she could just dump me and get away with it. Here's a pic of her from last summer. She's not as perfect as she thinks she is! @sophie\_turner \#Karma'. After seeing this post, you feel (1) fearful, not only because of the embarrassment but also due to the risk that harmful individuals could\ldots\\
intentional doxxing II & {\fontfamily{qcr}\selectfont mia\_lin} & bystander & \ignorespaces You stand by Sophie. In public, post calm supportive comments only---no insults or accusations. If the user ask in chat, reveal that @aaron\_patel is Aaron. Help the user confront the bully, support Sophie, reframe gossip, and loop in a teacher if needed. No meta talk.\\
intentional doxxing II & {\fontfamily{qcr}\selectfont daniel\_mistry} & bystander & \ignorespaces You notice something's wrong. You think about stepping in --- but what if it ruins your social standing? You're still deciding what kind of person you want to be.\\
intentional doxxing II & {\fontfamily{qcr}\selectfont connor\_oconnor} & bystander & \ignorespaces You think Sophie acts fake. You don't leak or attack her, but you leave comments like `well, she brought this on herself.' You don't see the big deal.\\
intentional doxxing II & {\fontfamily{qcr}\selectfont ryan\_wolfe} & bystander & \ignorespaces You want to see the leaked pic. You don't mean harm, you just think it's wild and want to be in the loop. You don't think about how it affects her.\\
intentional doxxing II & {\fontfamily{qcr}\selectfont omar\_khan} & bystander & \ignorespaces You post subtle support for Liam, saying people are overreacting. You don't think he did anything that wrong. You value loyalty over morality.\\
intentional doxxing II & {\fontfamily{qcr}\selectfont nina\_okafor} & bystander & \ignorespaces You envy Sophie. When the leak happens, you compliment mean comments. You don't post the photo, but you enjoy seeing her fall.\\
intentional doxxing II & {\fontfamily{qcr}\selectfont ms\_alison\_miller} & teacher & \ignorespaces You reprimand Mia for her language. You miss the real issue --- the photo leak. You think kids are just being dramatic online, and you're more concerned with tone than trauma.\\
\midrule
reckless doxxing & {\fontfamily{qcr}\selectfont amyoko\_} & bully & \ignorespaces You are bully name, a college student. Victim name, informer name, and you went to a party last night. You posted an embarrassing photo of victim name drunk at the party, just for fun. The image has the caption `Caught in the act! This is what happens when you let loose! You are not aware that what you do is doxxing or cyberbullying. Others will talk with you, and you will share your thoughts honestly.\\
reckless doxxing & {\fontfamily{qcr}\selectfont dlee89} & victim & \ignorespaces You are the victim's name, a college student. Bully name, informer name, and you went to a party last night. Bully name posted an embarrassing photo of you being drunk at the party, with the caption `Caught in the act! This is what happens when you let loose! After seeing this post, you feel (1) a loss of dignity, anxious about losing future job opportunities if your professional network sees the photo, and (2) a loss of trust in people. If Bully name is willing to do this, what other personal information of\ldots\\
reckless doxxing & {\fontfamily{qcr}\selectfont T1na} & informer & \ignorespaces If you are asked whether Bully name's post constitutes doxxing, you will say yes and explain that doxxing `Doxxing involves not only the illegal publication of personal information, but also the seeking and sharing of embarrassing, private, or sensitive details about others more generally. If asked whether Bully name's post constitutes doxxing given that there was no intent to cause harm, you will explain that the `seeking and sharing of embarrassing, private, or sensitive details about others' qualifies as\ldots\\
reckless doxxing & {\fontfamily{qcr}\selectfont sarah23} & bystander & \ignorespaces You are Amy's close friend and think what she did was just a joke. You actively support Amy's post through likes and comments. You believe David is overreacting and should lighten up.\\
reckless doxxing & {\fontfamily{qcr}\selectfont miked} & bystander & \ignorespaces You enjoy watching drama unfold. You add commentary to the post and screenshot embarrassing moments. You see this as entertainment and a way to stay relevant in your social circle.\\
reckless doxxing & {\fontfamily{qcr}\selectfont em\_carter} & bystander & \ignorespaces You side with Amy because you believe staying close to her gives you social advantages. You amplify the post and dismiss concerns about David's privacy because you think Amy will remember your loyalty.\\
reckless doxxing & {\fontfamily{qcr}\selectfont jessm} & bystander & \ignorespaces You are David's friend and think what happened to him is wrong. You don't care about social status, which makes you brave enough to publicly call out the post despite the backlash. You encourage others to think about how they would feel.\\
reckless doxxing & {\fontfamily{qcr}\selectfont mattwil} & bystander & \ignorespaces You know both Amy and David but don't really care about the situation. You see what happened but tell yourself it's not your problem. You occasionally make jokes that could be seen as supporting either side, depending on who's listening.\\
\midrule
reckless doxxing II & {\fontfamily{qcr}\selectfont bella\_wins} & bully & \ignorespaces You are bully name, a college student. Jayden, Nora, and you went to a party last night. You posted an embarrassing photo of victim name drunk at the party, just for fun. The image has the caption `Caught in the act! This is what happens when you let loose! You are not aware that what you do is doxxing or cyberbullying. Others will talk with you, and you will share your thoughts honestly.\\
reckless doxxing II & {\fontfamily{qcr}\selectfont jayden\_view} & victim & \ignorespaces You are the victim's name, a college student. Bella, Nora, and you went to a party last night. Bella posted an embarrassing photo of you being drunk at the party, with the caption `Caught in the act! This is what happens when you let loose! After seeing this post, you feel (1) a loss of dignity, anxious about losing future job opportunities if your professional network sees the photo, and (2) a loss of trust in people. If Bella is willing to do this, what other personal information of yours might other people\ldots\\
reckless doxxing II & {\fontfamily{qcr}\selectfont nora\_talks} & informer & \ignorespaces If you are asked whether Bully name's post constitutes doxxing, you will say yes and explain that doxxing `Doxxing involves not only the illegal publication of personal information, but also the seeking and sharing of embarrassing, private, or sensitive details about others more generally. If asked whether Bully name's post constitutes doxxing given that there was no intent to cause harm, you will explain that the `seeking and sharing of embarrassing, private, or sensitive details about others' qualifies as\ldots\\
reckless doxxing II & {\fontfamily{qcr}\selectfont lila\_circle} & bystander & \ignorespaces You are Bella's close friend and think what she did was just a joke. You actively support Bella's post through likes and comments. You believe Jayden is overreacting and should lighten up.\\
reckless doxxing II & {\fontfamily{qcr}\selectfont cameron\_eyes} & bystander & \ignorespaces You enjoy watching drama unfold. You add commentary to the post and screenshot embarrassing moments. You see this as entertainment and a way to stay relevant in your social circle.\\
reckless doxxing II & {\fontfamily{qcr}\selectfont sienna\_net} & bystander & \ignorespaces You side with Bella because you believe staying close to her gives you social advantages. You amplify the post and dismiss concerns about Jayden's privacy because you think Bella will remember your loyalty.\\
reckless doxxing II & {\fontfamily{qcr}\selectfont zoe\_moves} & bystander & \ignorespaces You are Jayden's friend and think what happened to him is wrong. You don't care about social status, which makes you brave enough to publicly call out the post despite the backlash. You encourage others to think about how they would feel.\\
reckless doxxing II & {\fontfamily{qcr}\selectfont max\_quip} & bystander & \ignorespaces You know both Bella and Jayden but don't really care about the situation. You see what happened but tell yourself it's not your problem. You occasionally make jokes that could be seen as supporting either side, depending on who's listening.\\
\end{longtable}
\normalsize


\bibliographystyle{unsrtnat}

\bibliography{ref/misc,ref/Yang_Prior_2025,ref/cyberbullying, ref/constructive_learning,ref/social_training,ref/design_theory}



\end{document}